\documentstyle[12pt,epsf]{article}
\input epsf
\def\beq{\begin{eqnarray}}   
\def\eeq{\end{eqnarray}}
\def\M4{\overline M}
\def\R4{\overline R}
\def\d{\delta^{(N)}(y_m)}

\begin{document}
\begin{flushright}
NYU-TH/00/08/08 \\
November 5, 2000
\end{flushright}

\vspace{0.1in}
\begin{center}
\bigskip\bigskip
{\large \bf Gravity on a Brane in Infinite-Volume  Extra Space}

\vspace{0.5in}      

{Gia Dvali, Gregory Gabadadze\footnote{Present address:
Theoretical Physics Institute, Univ. of Minnesota, Minneapolis, MN 55455}}
\vspace{0.1in}

{\baselineskip=14pt \it 
Department of Physics, New York University,  
New York, NY 10003 } 
\vspace{0.2in}
\end{center}

\vspace{0.9cm}
\begin{center}
{\bf Abstract}
\end{center} 
\vspace{0.1in}

We generalize the mechanism proposed in  [hep-th/0005016] and 
show that a four-dimensional relativistic tensor 
theory of gravitation can be obtained on a delta-function brane 
in flat infinite-volume  extra space. 
In particular, we demonstrate that the induced Ricci scalar 
gives rise to Einstein's gravity on a delta-function 
type brane if the number of space-time dimensions 
is bigger than five. The bulk space exhibits the 
phenomenon of infrared transparency. That is to say, the bulk  
can be probed  by gravitons with vanishing four-dimensional 
momentum square, while it is unaccessible  to higher modes.
This provides an attractive framework for solving 
the cosmological constant problem.

\newpage

\section{Introduction and summary}

The  size of compact extra dimensions 
could be as big as a millimeter if there is a brane in extra 
space on which Standard Model particles are 
localized \cite {ADD}. Gravity, in this framework,
spreads into the large compact extra dimensions, 
and that is why it is weak compared to other 
interactions.

Yet another possibility is to maintain a brane-world
hypothesis without actually compactifying extra space, but rather 
giving it some nonzero curvature \cite {RandallSundrum}. 
As a result,  the invariant volume of the extra space is still finite 
and is determined by the bulk cosmological constant 
\cite {RandallSundrum}. 

The aim of the present work is to look for 
models in which the extra dimensions are uncompactified 
and, moreover, they are flat at infinity. 
In this case,  the invariant volume  of extra space is 
truly infinite. 
The matter and gauge fields can be localized on such a brane 
using the field-theoretic mechanisms of Refs. \cite {Rubakov,Akama}
and \cite{GMLoc} respectively, 
or the string theory mechanism for localization
of matter and gauge fields on a D-brane \cite {Polchinski}.
The immediate challenge, however,  is to obtain 
four-dimensional gravity on the brane worldvolume
where the Standard Model particles live. 

In Ref. \cite {DGPind}  
the mechanism was proposed by which four-dimensional 
gravity can be obtained on a brane 
in 5-dimensional flat space-time.
The mechanism is based on the observation  that the 
localized matter fields on a brane (which 
couple to bulk gravitons) 
can generate  the localized four-dimensional 
worldvolume kinetic term for gravitons. 
That is to say, four-dimensional gravity is ``pulled over'' 
(or induced) from the bulk gravity  
to  the brane worldvolume by the matter fields confined to 
a brane.

It was shown in Ref. \cite {DGPind} that the four-dimensional Newton
law is recovered on a brane. On the other hand, the relativistic effects
of this theory differ from those of Einstein's 
gravity. In fact, the induced  gravity on a brane in 5-dimensional 
Minkowski space is tensor-scalar gravity \cite {DGPind}. 

In this work we would like to 
pursue a more general strategy 
and study the mechanism of Ref. \cite {DGPind}  
on a brane in space-time with the number of 
dimensions bigger than five. 
The main  motivation 
to go beyond 5-dimensions comes from the 
well known fact that in 5-dimensional space  there is 
no {\it static (non-inflating)} and {\it stationary} 
domain wall solution with a  nonzero positive tension 
\cite {Vilenkin,IpserSikivie,NemanjaLinde} 
which could be embedded in 
a flat and infinite transverse space. 
In contrast, in dimensions bigger than five, such 
nonzero tension solutions with a non-inflating 
worldvolume can be found. 
   
In the present work  we will show  
that using the mechanism of Ref. \cite {DGPind} 
one can obtain the four-dimensional laws of 
{\it Einstein's gravity} on a {\it delta-function type} 
brane world-volume if $D>5$.  
This difference between 
five-dimensional and $D>5$-dimensional theories 
can be traced back to the fact
that in the former case the transverse to the brane space  
is one-dimensional and the transverse Green  functions 
are finite at the origin, while in the latter case 
these functions diverge at zero.

The physical picture which we obtain can be 
summarized as follows.
There is a flat non-zero tension 
3-brane embedded in $D>5$ space-time which is also flat 
at infinity.  
Thus, the invariant volume of the extra space is infinite. 
The localized matter on the brane couples to
the bulk gravity. As a result of this coupling 
the four-dimensional kinetic term for gravitons is generated 
via  quantum loops on the worldvolume.
These ``four-dimensional'' gravitons
are nothing but part of bulk gravitons. Furthermore,  
the induced four-dimensional kinetic term  gives rise 
to four-dimensional laws of gravity on a brane worldvolume.

As a next step  we would like  to  address a  more 
pragmatic question:
The infinite-volume theories -- what are they good for?
Certainly they have an independent academic interest. 
Furthermore, they could lead 
to the modification of the laws of gravity 
at ultra-large distances \cite {Kogan,GRS,Csaki1,DGP1,WittenCC}.
Most importantly, as it was pointed out in Refs. \cite {DGP1,WittenCC},
these theories give a new, yet unexplored  way of 
thinking about  the cosmological constant problem. 
This will be discussed in details in 
Section 6. Here, we just briefly reiterate the arguments of
\cite {DGP1,WittenCC}. One could start with
a supersymmetric theory in high-dimensions,
let us say with superstring theory, M-theory    
or some of the low-energy supergravity
truncations. If the brane on which we live is a non-BPS brane,
then it can be used to break all the supersymmtries
on the worldvolume theory \cite {GiaMisha} (stability 
of such a brane can in principle be warranted  
by topological reasons, or equivalently by giving 
some conserved charges to the brane). 
Thus, classically the brane worldvolume is not supersymmetric, 
while the bulk is supersymmetric. The question is 
whether the bulk supersymmetry could be preserved 
in the full quantum theory. The answer is positive and the key point here
is that the bulk has an infinite volume. Because of this, 
transmission of SUSY breaking from the brane worldvolume to the 
bulk vanishes as inverse volume. 
Thus, one has SUSY in the whole high-dimensional 
theory without Fermi-Bose degeneracy in 4D theory\footnote{This is close
in spirit to  the $(2+1)$-dimensional
example of \cite {Witten3D} and $(3+1)$-dimensional example
of \cite{GiaMon}, although the mechanism
and the properties are very different.}.  
Furthermore, imposing the 
condition that the bulk preserves the  R-symmetry along with local 
SUSY, we obtain that the bulk cosmological constant is zero to all orders 
in full quantum theory. 
How about the 4D cosmological constant which we should supposedly 
be observing? There are two parts to the  answer to  this question.
The first one is that 
the 4D cosmological constant which is produced by 
the brane worldvolume theory can be  re-absorbed by
rescaling of the brane tension. Such a brane, in $D>5$ will or 
will not  inflate
(in $D=5$ it inflates with necessity \cite {Vilenkin,IpserSikivie,
NemanjaLinde}). The reason why it might  not inflate in the 
present case is that the bulk supersymmetry and the induced term 
might not allow it  to do so.
Seemingly alternative, but in fact an equivalent
way of thinking about this is as follows: 
At extremely  low energies,
the theory at hand is not four-dimensional. Rather it is 
higher-dimensional due to the presence of the infinite-volume bulk
(in this sense the physics is inverted upside down
compared to  the theories with compact of warped 
extra dimensions, where at lower energies the  theory becomes more 
and more four-dimensional). 
Therefore, what one ``sees'' at extremely low energies, 
is not the 4D cosmological cosmological constant but 
rather the higher-dimensional
cosmological constant which is zero due to bulk SUSY 
and R-symmetry. We will discuss these  issues in details in section 6.

\section{Gravity on a brane in extra Minkowski space}

Let us suppose that the bulk action has the 
following general form:
\beq
S_{\rm bulk }~=~\int d^{4+N}X~\sqrt{|G|}~{\cal L}\left 
(G_{AB}, ~{\cal R }_{ABCD},
~\Phi  \right )~,
\label{act1}
\eeq
where the capital Latin indices run over $D=(4+N)$-dimensional space-time. 
$G_{AB}$ denotes the metric of D-dimensional 
space-time, ${\cal R}_{ABCD}$ is the D-dimensional 
Riemann tensor, and  $\Phi$ 
collectively denotes all other bulk fields. 
Suppose that there is a 
3-brane embedded in this space.  
The surface term which is necessary
for the correct variational procedure 
for the action  will be implied in the bulk action 
in all the discussions below. 
We simplify further our presentation 
by considering the whole picture in the 
effective field theory framework. Our goal will be to find
the properties of the system which are independent of 
these simplifications.
Let us suppose that the 3-brane is localized in the extra 
space so that it asymptotes to a flat space at infinity.
We split the coordinates in D-dimensions as follows:
\beq
X^A~=~(x^{\mu},~y^m)~,
\eeq
where Greek indices run over four-dimensional brane worldvolume,
$\mu=0,1,2,3$~, and small Latin 
indices over the space transverse to a 
brane, $m,n,i,j=4,5,..,4+N$. 

The Dirac-Nambu-Goto action for a brane takes the form:
\beq
S_{\rm 3-brane}~=~-T 
~\int d^{4}x~\sqrt{|{\rm det} {\bar g}|}~,
\label{brane0}
\eeq 
where $T$ stands for the brane tension and 
${\bar g_{\mu\nu}}=\partial_\mu X^A\partial_\nu X^B G_{AB}$ 
denotes  the induced metric on a brane.
In the most part of this  work, unless stated otherwise,
we treat a  brane as a 
{\it delta-function type} singular source which is located at
a point $y_m=0$ in the extra space. Moreover, for the moment  
we neglect its  fluctuations. Therefore, the  induced metric 
can be written  as follows:
\beq
{\bar g}_{\mu\nu}~(x)~=~G_{\mu\nu}\left (x, y_n=0 \right )~.
\label{gind}
\eeq
In general, there could  be localized matter fields on the brane 
worldvolume. These can be taken into account by writing the 
following action for the brane:
\beq
{\tilde S}^{\rm matter}_{\rm 3-brane}~=~S_{\rm 3-brane}~+~
\int d^{4}x~\sqrt{|{\rm det} {\bar g}|}~{\tilde {\cal L}} (\phi)~,
\label{brane1}
\eeq  
where $\phi$ denotes collectively all the localized 
fields for which the four-dimensional 
Lagrangian density is ${\tilde {\cal L}}$. 

Note that in  the 
classical theory, which we are discussing so far, the
4D Ricci scalar on the brane worldvolume is not present. 
Thus, the localized particles separated at a distance $r$
on a brane interact via  the $(4+N)$-dimensional
gravitational force-law, that is $F\sim 1/r^{2+N}$.
This holds as long as the classical theory is concerned. 
However, in full quantum theory the 4D Ricci scalar
will be generated (along with other terms) on a brane worldvolume. 
This is due to quantum loops  
of the matter fields which are {\it localized} on a brane worldvolume
\cite {DGPind} (see also Appendix A). 
As a result, the following   worldvolume terms should be included into  
the consideration in the full quantum theory:
\beq
S_{\rm ind}~=~{\M4}^2~\int d^{4}x~
\sqrt{|{\rm det}{\bar g}|}~\left [ {\bar \Lambda}~+~ {\R4}(x)~+{\cal O}
\left ( {\R4}^2  \right ) \right ]~,
\label{4DR}
\eeq  
where ${\M4}$ is some parameter which depends 
on details of the worldvolume model \cite {DGPind} (see also 
Appendix A). 
For phenomenological reasons $\M4$ should be of the order
of the 4D Planck scale:
\beq
\M4~\sim M_{\rm Pl}~\simeq~10^{19}~{\rm GeV}~.
\eeq 
In section 5 we will discuss how this big scale could  be 
generated.  

After the brane is included in the theory 
translation invariance in the directions transverse to the 
brane is broken. On the other hand, four-dimensional 
reparametrization invariance 
(``gauge invariance'') along the brane world-volume 
coordinates is preserved in the theory.
This invariance requires  that 
the induced terms in (\ref {4DR}) preserve this 4D gauge symmetry.
Possible operators with derivatives with respect to  
the transverse coordinates  
are not induced on the worldvolume of a delta-function type brane
since the localized particles on a brane have zero transverse momenta. 
This can also be seen by explicit calculations of loop diagrams
in gauge invariant regularization schemes.

$\bar \Lambda$  in (\ref {4DR}) is an 
induced four-dimensional cosmological 
constant. The role of this term is that it renormalizes 
the brane tension.  Furthermore, ${\R4}(x)$
is the four-dimensional Ricci scalar which is constructed 
out of the induced metric ${\bar g}_{\mu\nu}(x)$ 
defined in (\ref {gind}). 

All  terms in  (\ref {4DR}) are  consistent with the 
symmetries of the theory in which conformal invariance and 
SUSY are broken in the  brane worldvolume.
Any realistic brane-world 
model should possess  these  properties\footnote{
Note  that the induced 4D cosmological term ${\bar \Lambda}$ is not 
generated if  the brane worldvolume theory is supersymmetric.
However, the other terms in (\ref {4DR}) 
will still be induced  in a non-conformal supersymmetric theory.}. 
Moreover, 
the terms  in (\ref {4DR})
are relevant operators of the four-dimensional
worldvolume theory which will be induced on the brane even if 
they were not present at the first place \cite {DGPind}.  
Since these terms are unavoidable in any realistic brane world scenario,
we should  study the physical consequences of term.  

The total  brane worldvolume action takes the form:
\beq
S_{\rm W}~=~-T^\prime 
~\int d^{4}x~\sqrt{|{\bar g}|}~+~
{\M4}^2~\int d^{4}x~
\sqrt{|{\bar g}|}~\left [{\R4}(x)~+{\cal O}
\left ( {\R4}^2  \right )~+...\right ]~,
\label{worldv}
\eeq
where ${\bar g}\equiv {\rm det} {\bar g}$, and 
$T^\prime\equiv T-{\bar \Lambda}~\M4^2$   is the
renormalized brane tension  which absorbs the induced
four-dimensional cosmological constant.  Dots in this expression 
stand for other possible worldvolume matter fields and 
interactions which we will omit  below for simplicity. 
The field theory on a brane worldvolume is  
an effective field theory with a cutoff\footnote{This statement will be 
elucidated in details in section 5.}. 
In the effective field theory framework 
the higher derivative terms appearing in (\ref {worldv}) are 
suppressed by higher powers of $\M4$ 
and can be neglected in the leading approximation.

\section{Four-dimensional  gravity on a brane}

In this section we 
study the laws of gravity on a brane 
with the worldvolume action given by (\ref {worldv}). 
We  adopt  a simplest setup in which  
the  gravitational part of the 
bulk Lagrangian contains only the Einstein term
while the gravity on the  worldvolume is 
given by (\ref {worldv}):
\beq
M^{2+N}~\int d^{4+N}X ~\sqrt {|G|}~{\cal R}_{(4+N)}~+~
\M4^2 ~\int d^4x ~\sqrt {|{\bar g}|}~{\R4}~.
\label{1}
\eeq 
Here, $M$ denotes the Planck constant of the 
bulk theory. For the simplicity of calculations we
will temporarily neglect the brane tension term, i.e.,  we put 
$T^\prime=0$. This is not essential as long as one is dealing with a 
theory in dimensions higher than 5. In this case, 
there can exist 
brane solutions with {\it static} worldvolume 
which have a {\it non-zero} tension.
Therefore, non-zero $T$  can consistently be 
restored  back (see sections 5 and 6). 
However, the case $D=5$ is special. As is well known
\cite {Vilenkin,IpserSikivie,NemanjaLinde}, any stationary nonzero
{\it positive} tension brane inflates in 5D. 
Hence, it is not feasible in 5D to go from the theory with 
a zero-tension brane
to a viable model where the brane tension has  a 
positive finite value.
Moreover, it  was claimed  in Ref. \cite {Z1} that  
any constant-curvature bulk in 5D 
cannot control the brane cosmological constant. 
We will discuss  these issues
in more details in section 6. 
Before that,  we assume that  the number of space-time 
dimensions is greater than 6 (although, as we will show  
below,  some of our results will
also be applicable for the 6-dimensional case.).

\subsection{Newtonian gravity on a brane}

First  we study Newtonian potential
between two localized masses on a  brane. 
For this we can drop temporarily 
the tensor structure in the graviton propagator. 
Effectively, this is equivalent to 
the exchange of a massless scalar mode in the worldvolume theory
\footnote{Even for warped backgrounds  equations for 
scalars are similar to those of gravitons \cite {BorutGiga}.}. 
Thus, we define the prototype  Lagrangian for this scalar:
\beq
M^{2+N}~\int d^{4+N}X~\left (\partial_A\Phi(x,y) \right )^2~+~
\M4^2\int d^{4}x~\left (\partial_\mu\Phi (x, y=0)\right )^2~.
\label{scalar}
\eeq
Here, the first and the second terms are  respectively counterparts  
of  the bulk Ricci scalar ${\cal R}_{(4+N)}$  
and the induced 4D Ricci scalar $\R4$  in (\ref {1})
\footnote{Note that the scalar model considered here is just used
as a demonstration for more complicated model of gravity. 
If a scalar field theory  is  considered
independently by its own, then on a brane with a finite width 
one expects  additional induced terms containing derivatives with 
respect to $y$. 
However, these terms are not generated for scalars on the delta  
function type brane considered in this work 
since all the particles 
in the matter loops which induce 
worldvolume terms have exactly zero transverse momenta.}.

We are looking for  the distance dependence of 
interactions in a  4D worldvolume theory. For this 
we should find the corresponding retarded Green  function  
and calculate the potential. 
The equation for the Green  function  
takes the form:
\beq
\left ( M^{2+N} \partial_A\partial^A~+
~\M4^2~\delta^{(N)}(y_m)~ \partial_\mu\partial^\mu 
\right ) ~G_R(x,
y_m; 0,0)~=~-~\delta^{(4)}(x) \delta^{(N)}(y_m)~,
\label{green}
\eeq
where $G_R(x,y_m; 0,0)=0$ for $x_0<0$.
 
The potential at a distance $r$ along the 
brane worldvolume  is determined as follows:
\beq
V(r)~=~\int G_R\left 
(t,{\overrightarrow x},y_m=0; 0,0,0\right 
) dt~,
\label{pot}
\eeq
where $r\equiv\sqrt{x_1^2+x_2^2+x_3^2}$. 
To find a solution of (\ref {green}) let us turn to 
Fourier-transformed quantities with respect to 
the worldvolume four-coordinates $x_\mu$:
\beq
G_R(x,y_m; 0,0)~\equiv~\int ~ {d^4p\over (2\pi)^4}~e^{ipx} 
~{\tilde G}_R(p,y_m)~. 
\label{Fourie}
\eeq 
In  Euclidean space equation (\ref {green})
takes the form:
\beq
\left (~ M^{2+N} (p^2-\Delta_N)~+~\M4^2~ 
p^2 ~\delta^{(N)}(y_m) ~\right )~  
{\tilde G}_R(p,y_m)~=
~\delta^{(N)}(y_m)~. 
\label{mom}
\eeq
Here $p^2$ denotes the square of an Euclidean four-momentum,
$p^2=p_4^2+p_1^2+p_2^2+p_3^2$, 
and $\Delta_N$ stands for the Laplacian of the N-dimensional 
transverse space. 

We look for the solution of (\ref {mom}) in the
following form:
\beq
{\tilde G}_R(p,y_m)~=~D (p,y_m)~B(p),
\label{DD}
\eeq 
where $D (p,y_m)$ is defined as follows:
\beq
(p^2-\Delta_N)~D (p,y_m)~=~\d~.
\label{ED}
\eeq
$B(p)$ is some function to be determined\footnote{
There are some subtleties related to the fact that seemingly one has 
a product of two singular functions in (\ref {mom}) and (\ref {DD}).
To avoid this, one should work with 
the regularized expression: ${\tilde G}_R(p,y_m)~=~\lim_{\epsilon_m
\rightarrow 0}D (p,y_m+\epsilon_m)~B_{\epsilon}(p)$. 
A  careful analysis with this regularization shows that 
the formal derivation with singular functions 
presented below is valid.}.
Using this decomposition one 
finds:
\beq
{\tilde G}_R(p,y_m)~=~{D(p,y_m)\over M^{2+N}~+~\M4^2p^2~D(p,0)}~.
\label{G}
\eeq
For the case $D>5 $ the function  $D(p,y_m)|_{y_m=0}$ diverges
\footnote{This is another reason why the case $D=5$ is 
exceptional, here 
$D(p,y_m)|_{y_m=0}$ is finite, see \cite {DGPind} and Appendix B.}.
Therefore, the expression  for ${\tilde G}_R(p,y_m) $ has a jump
at $y_m=0$.  In the brane worldvolume, i.e., for
$y_m=0$, we find
\beq
{\tilde G}_R(p,y_m=0)|~=~{1\over \M4^2 p^2}~,
\eeq 
which is nothing but the Green  
function for a four-dimensional theory.

Therefore, we conclude that the static potential
between a couple of point-like sources on a brane
in $D>5$ scales with distance between them  as 
a four-dimensional potential:
\beq
V(r, y_m=0)~=~{1\over 8\pi \M4^2}~{1\over r}~.
\label{4DV}
\eeq 
In the approximation of a delta-function type 
brane,  which we adopt in the present work, this  
behavior is expected to hold all the way up to infinite distances.
However, for a finite thickness brane, the four-dimensional Newton law
might  be changed at very 
large (presumably of the order of the present Hubble size) distances,
see discussions in section 4.

Outside of the brane, i.e., for $y_n\neq 0$,  
there are two, physically distinct cases to consider. These 
differ by the value of the four-momentum square:
$p^2=0$,  and  $p^2\neq 0$. We will discuss them in turn.

\vspace{0.1in}

(i) {\it Vanishing four-momentum square}, $p^2=0$:  

\vspace{0.1in}

In this case the expression (\ref {mom}) reduces to the equation for 
Euclidean Green's function in the  transverse N-dimensional space. 
In general, this Green's function  is well known:
\beq
{\tilde G}_R(0,y\neq 0)~\propto \lim_{p^2\to 0} ~{1\over M^{2+N}}~
 \left ( {p\over |y|}  \right )^{N-2\over 2}~
K_{N-2\over 2} (p~|y|)~,
\label{green1}
\eeq
where $|y|\equiv \sqrt{y_1^2+y_2^2+...+y_N^2}~$ and $K(p|y|)$ denotes the 
McDonald function.
For $p^2=0$ this function scales as follows: 
\beq
\sim {1\over ~~~|y|^{N-2}}, ~~~~{\rm for}~~~N>2~.
\label{N}
\eeq
For the $D=6$ case ($N=2$) this has 
a  logarithmic singularity at $p^2=0$: 
\beq
\sim ~\ln(p|y|)|_{p^2=0} \to \infty,  ~~~~{\rm for}~~~N=2~.
\eeq

Therefore, for $p^2=0$ there exists a solution 
in the form  of the N-dimensional Green  function 
(\ref {green1}). This means  that the $p^2=0$  
mode provides interactions between worldvolume states and bulk states. 
This mode should be produced with a non-zero three-momentum 
along the brane worldvolume.

\vspace{0.1in}

(ii) {\it Non-vanishing four-momentum square}, $p^2\neq 0$:  

\vspace{0.1in}

In this case physics  is rather different. 
Outside of a brane, i.e., for  $y_m\neq 0$,  the Green's function
vanishes:   
\beq
{\tilde G}_R(p,y_m\neq 0)|_{p^2\neq 0}~=~0~.
\eeq
In other words,  the $p^2\neq 0$ mode  
cannot give rise to any interactions of the localized worldvolume 
matter  with  the matter which is placed in the bulk. 
Therefore, the interaction between a particle 
localized on a brane and a particle placed in the bulk
is completely determined by the $p^2=0$ solution and has the form
given in Eq. (\ref {N}).

\subsection{Tensor structure of the graviton propagator}

We have established in the previous subsection that the 4D Newton law 
is reproduced on the brane. However, this is not enough. 
The relativistic 
theory of gravitation  on the brane should be the Einstein 
tensor gravity plus possible higher derivative terms. 
In the minimal 5D setup  studied
in Ref. \cite {DGPind} this is not the case: the relativistic
model on a brane worldvolume is  tensor-scalar gravity.
The unwanted additional scalar there  comes from the 
extra polarization of a bulk 5D graviton \cite {DGPind}.  
In this section we will show that in $D>5$ 
the four-dimensional theory on a delta-function type 
brane is  consistent 4D
tensor gravity.
 
To show this we need to study 
the tensor structure of the graviton propagator. 
Let us   introduce the metric fluctuations:
\beq
G_{AB} ~=~ \eta_{AB} ~+~h_{AB}~.
\eeq
We choose {\it harmonic~ gauge} in the bulk: 
\beq
\partial^A h_{AB}~ =~{1\over 2}~ \partial_B h^C_C.
\label{gauge}
\eeq 
It can be checked that the choice
\beq
h_{\mu m} ~=~0,~~~m=4,...,4+N~, 
\label{mu5}
\eeq
is consistent with the equations of motion for the action 
(\ref {1}) which is amended by a point-like source term  located on a brane.
Thus, the surviving components of $h_{AB}$ are 
$h_{\mu\nu}$ and $h_{mn}$.
In this gauge the $\{mn\}$ components of Einstein's  
equations yield:
\beq
(6-D)~ \partial_A\partial^A ~h^m_m ~=~
(D-4)~\partial_A\partial^A~ h^\mu_\mu~.
\label{mn}
\eeq
Indices  in all these equations 
are raised and lowered  by a flat space  metric tensor.
Finally, we come to the $\{\mu\nu \}$ components
of the Einstein equation. After some 
rearrangements  these    take  the form:
\beq
M^{2+N} \left ( \partial_A\partial^A~h_{\mu\nu}
-{1\over 2} \eta_{\mu\nu} \partial_A\partial^A h^\alpha_\alpha
-{1\over 2} \eta_{\mu\nu} \partial_A\partial^A h^n_n \right )~+~
\nonumber \\
\M4^2~\d~ \left ( 
\partial_\alpha \partial^\alpha~h_{\mu\nu}
-{1\over 2} \eta_{\mu\nu} \partial_\beta\partial^\beta h^\alpha_\alpha
+{1\over 2} \eta_{\mu\nu} \partial_\beta\partial^\beta h^n_n
-\partial_\mu \partial_\nu h^n_n 
\right )~=\nonumber \\ 
~T_{\mu\nu}(x) ~\d~.
\label{basic}
\eeq
Here, the energy-momentum tensor for the localized on a brane 
source is denoted by $T_{\mu\nu}(x) ~\d~$.
As before, there are two groups of terms on the l.h.s. of the equation
(placed in separate  parenthesis).
The first group originates from  the 
bulk Ricci scalar ${\cal R}_{(4+N)}$ and the second  one 
from  the induced 4D Ricci  scalar $\R4$. 

In order to determine the tensor structure of the 
graviton propagator it is convenient to bring this expression to the 
following form:
\beq
\left (  M^{2+N} ~\partial_A\partial^A+\M4^2\d
\partial_\alpha \partial^\alpha \right )h_{\mu\nu}=\left \{  
T_{\mu\nu}(x) -{1\over 2}\eta_{\mu\nu}T^\alpha_\alpha(x)
  \right \}\d \nonumber \\ 
- M^{2+N}
{1\over 2} \eta_{\mu\nu} \partial_A\partial^A h^n_n~+
~\M4^2~\d \partial_\mu \partial_\nu h^n_n~. 
\label{mten}
\eeq 
The tensor structure of the terms 
which contain $T_{\mu\nu}$ on the r.h.s. of this equation
is that of for-dimensional Einstein's
gravity:
\beq
~\left \{  
T_{\mu\nu}(x) ~-~{1\over 2}\eta_{\mu\nu}~T^\alpha_\alpha(x)
  \right \}~.
\label{T}
\eeq
However, as we see there are  two  additional 
term on th r.h.s. of (\ref {mten}) 
(written  in the second line of  (\ref {mten})).
Let us start with the second term. In the momentum space this
term  is proportional to the product $p_\mu p_\nu$. Therefore,
its contribution vanishes when it is convoluted with a conserved 
energy-momentum tensor. In that respect,  it is similar to 
gauge dependent terms occurring in graviton 
propagators. This term  is harmless. 
The first term, $\eta_{\mu\nu} \partial_A\partial^A h^n_n$, however, 
cannot be removed by gauge transformations. Depending on
dimensionality of space-time, this term  
might or might not give rise to additional contributions
to 4D  gravity.
For instance, in accordance with (\ref {mn}), in the four-dimensional case
($D=4$)  the expression
$\eta_{\mu\nu} \partial_A\partial^A h^n_n$ vanishes. Therefore, 
we recover  ordinary 4D Einstein's  gravity. 
However, in $D=5$, as was  shown in \cite {DGPind}, 
this extra term gives rise to the additional scalar exchange
and, as a result,  the 4D  theory  is  tensor-scalar
gravity. Our goal is to establish 
which of these possibilities is realized
in higher dimensions. 

As before,  let us make the Fourier-transform with respect to 
the four worldvolume coordinates.  Equation (\ref {mten})
can be rewritten in the following form:
\beq
\left [ M^{2+N}~ (p^2~-~\Delta_N)~+~\M4^2 ~\d~ p^2 \right ]
{\tilde h}_{\mu\nu}(p, y_m)~{\tilde T}^{\prime \mu\nu}~=~
\nonumber \\
~\left \{  
{\tilde T}_{\mu\nu}{\tilde T}^{\prime \mu\nu} ~-~
{1\over 2}~{\tilde T}^\alpha_\alpha 
{\tilde T}^{\prime \beta}_\beta
\right \}~\d~-~{1\over 2}{\tilde T}^{\prime \alpha}_\alpha (p^2-\Delta_N)
{\tilde h}^m_m~. 
\label{mom1}
\eeq
Here, the sign tilde denotes the Fourier-transformed
quantities and ${\tilde T}^{\prime\mu\nu}$ is a  conserved 
energy-momentum tensor in  the momentum space. 
As in the previous subsection there are two different types of solutions
to this equation. Let us  start with the solution on the 
brane worldvolume, i.e., that with $y_n=0$.
Using Eqs. (\ref {mn}), (\ref {mom1}) and the methods we used for the 
scalar case (the decompositions similar to (\ref {DD})) we find that:
\beq
{\tilde h}_{\mu\nu}
(p, y_m)~{\tilde T}^{\prime \mu\nu}|_{y=0}~
=~\left [ {\tilde T}_{\mu\nu}{\tilde T}^
{\prime \mu\nu} ~-~
{1\over 2}~{\tilde T}^\alpha_\alpha 
{\tilde T}^{\prime \beta}_\beta  \right ]~{1\over \M4^2 p^2}~. 
\label{4DT}
\eeq
This is a perfectly 4D solution. Therefore, we 
find that the distance dependence 
and the tensor structure of 
the graviton propagator on a delta-function type brane in $D>5$  
is that of 4D Einstein's gravity. 

Having this result obtained, let us investigate what happens outside 
of the brane, i.e., for $|y|\neq 0$.
As with scalars in the previous subsection,  
there are two different  physical cases  to consider.   

\vspace{0.1in}

(i) {\it Vanishing four-momentum square}, $p^2=0$: 

\vspace{0.1in}

After some algebra one finds the following solution:
\beq
{\tilde h}_{\mu\nu}
(0, y_m)~{\tilde T}^{\prime \mu\nu}~|_{p^2=0}=
~\left [ {\tilde T}_{\mu\nu}{\tilde T}^
{\prime \mu\nu} ~-~
{1\over D-2}~{\tilde T}^\alpha_\alpha 
{\tilde T}^{\prime \beta}_\beta  \right ]~D(0,y)~.
\label{DD1}
\eeq
The expression for $D(0,y)$ is given in (\ref {N}). 
This is just a  N-dimensional solution with a 
D-dimensional tensor structure.   
Furthermore, this solution 
gives rise to interactions (\ref {DD1})  between the matter 
which is localized
on the brane and the matter which is placed in the bulk.  
This mode should  be produced with a nonzero three-momentum
directed along the brane worldvolume. 

\vspace{0.1in}

(ii) {\it Non-vanishing four-momentum square}, $p^2\neq 0$:

\vspace{0.1in}

As is expected, this case is similar to that  with scalars.   
Here we find: 
\beq
{\tilde h}_{\mu\nu}
(p, y_n)~{\tilde T}^{\prime \mu\nu}|_{y\neq 0}~=~0.
\label{4D}
\eeq
Thus, the $p^2\neq 0$ mode  cannot produce interactions between
the brane worldvolume matter and the bulk matter.

Let us note that in the worldvolume theory 
there are remaining components of the high-dimensional
metric, $h_{mn}$, which from the four-dimensional
standpoint look as scalar particles
with no mass gap. In the full quantum theory there is 
no symmetry which protects masses
of these states against radiative corrections. 
Therefore, they are expected to acquire on the brane 
potentials (masses) of 
order the cutoff of the worldvolume model. 

Summarizing this section we conclude that
the four-dimensional gravity on the delta-function type brane 
with the induced 4D Ricci scalar  
(in $D>5$ case) is consistent  tensor gravity with correct
Newtonian potential and the correct Einstein  tensor structure of 
the graviton propagator.

\section{Infrared transparency of extra space}

The bulk of this  paper is devoted to the study of a delta-function type 
brane. A similar consideration for a ``fat'' brane,
which turns out to be much more involved and rich,
will be given elsewhere \cite {GGnew}. In the present and next sections,
however, we will discuss some qualitative features 
which should emerge when ``fat'' branes are considered. 

In higher-dimensional theories one 
usually expects that extra space  
can be probed  in very high  energy accelerator experiments or,
in  very short distance gravity measurements.
This is certainly true when 
there is no induced kinetic terms on the brane.  
In the scenarios where  the extra space is compact,
or, alternatively, if it is warped 
as in  \cite {RandallSundrum}
the invariant 
physical volume of the extra space is finite. 
In the former case this is 
true by definition,  while in the case of warped 
but non-compact spaces this is also true since 
there is a nontrivial warp-factor which makes 
the invariant volume finite $\int_{-\infty}^{+\infty}dy 
\sqrt{g} <\infty$ (although the proper distance 
in this case is infinite). 
Therefore, at very high energies (or equivalently short distances)
the extra finite-volume space is probed. 

The natural question to ask is whether the same phenomenon 
holds in infinite-volume theories discussed in the present work.
We will argue below that physics of the infinite-volume
theories is inverted upside down 
compared to the finite-volume theories mentioned above. 
In fact, we will  argue that the extra bulk space might be probed
only at extremely small (close to zero) energies, or equivalently,
at ultra-large (close to the present Hubble size) distances.

To see this let us recall the results of the previous sections. 
There we found that the Green function on a brane 
behaves precisely as a 4D solution.  
That is to say, the worldvolume laws of gravity 
are  four-dimensional up to infinite distances.
We also have found that the solution  with 
zero four-dimensional momentum 
square, $p^2=0$, has the dependence on 
the bulk coordinates.
On the other hand, any other solution 
with $p^2\neq 0$ is zero in the bulk.
Thus, we say that
the bulk is transparent only for the $p^2=0$ mode. 
This phenomenon holds as long as 
the brane is a delta-function type source. 

However, the brane at hand might have certain 
finite transverse width. 
This could be set  by the size of the core of the 
brane if it appears as a smooth soliton in the bulk, or 
by  the effective size of the transverse fluctuations of the brane.  
In any case, if the brane has a finite width, there 
could  exist the modes with tiny but still nonzero $p^2$ which 
would  be able to leak into the bulk.  
In this respect, one could  define  
a certain critical momentum,
let us call it $p_c$,  below  which
the theory could become  higher-dimensional. Above this momentum the 
worldvolume theory behaves as a 4D model. Higher we go in momenta 
farther we are from $(4+N)$-dimensional theory, and the worldvolume
physics becomes more-and more four-dimensional. 
This is precisely opposite to what 
one obtains in theories with finite-volume extra dimensions\footnote{
Note that due to the specific nature of the $D=5$ 
case  this phenomenon takes place  in 5D models even for a brane which has 
a zero width (a delta-function
type brane) \cite {Kogan,GRS,Csaki1,DGP1}.}. 

In terms of distances, we define the critical distance 
$$
r_c~\equiv~{1\over p_c}~.
$$    
Below this distance physics is four-dimensional. However, for distances
bigger  than $r_c$ gravity changes.
On phenomenological grounds, $r_c$ should 
at least be of the average size of clusters of galaxies or so. 
The shorter distances we probe,
physics becomes  more and more four-dimensional.

Let us try to make these arguments a bit more quantitative.
For this let us look at   the plane wave solutions of
equations of motion (for simplicity we drop
the tensor structures again):
\beq
\left (~ M^{2+N} (p^2-\Delta_N)~+~\M4^2~ 
p^2 ~\delta^{(N)}(y_m) ~\right )~  
e^{ipx}f(y)~=~0.
\label{plane}
\eeq
There is only one plane-wave  solution to this equation
with $f(0)\neq 0$,
that is the wave with $p^2=0,~~f(y)={\rm const.}$.
The reason for such a behavior is that the term with $\d$ 
in (\ref {plane}) dominates over the first term if $p^2\neq 0$ and  
$D>5$. Thus, the resulting intrinsic 4D nature of the theory for $p^2\neq 0$.
On the other hand, for  $p^2=0$ the second term in (\ref {plane}) 
can be made small and  physics is determined  by the 
first term which naturally gives rise to D-dimensional results
\footnote{This suppression is similar to the one found
in Ref. \cite {DGPph}.}.

Let us now  suppose that 
the brane has a very small but finite width $B$. 
In this case, the $\d$ function in (\ref {plane}) should be 
replaced by some smooth spread  function. 
As before,  there are two competing terms in  (\ref {plane}),
one is D-dimensional and another one is 4-dimensional. 
Qualitatively, the effects of these terms in the  
vicinity of a brane are 
weighted respectively  by the quantities: 
\beq
M^{2+N}~ (p^2-{1\over B^2}) ~~~{\rm and}~~~~\M4^2~ 
p^2 {1\over B^N}~.
\label{two}
\eeq
At large momenta and $N>2$ the second term dominates since $B$ is very 
small. Therefore, 
the theory is 4-dimensional in that regime.  However, as we discussed above,
there is a critical value of the momentum, $p_c$, at which the 
two terms in (\ref {two}) could  become comparable. 
This value can be estimated:
\beq
p_c^2~\sim ~{ M^{2+N}~ B^{N-2} \over M^{2+N}~B^{N}~+~\M4^2}~.
\label{pc}
\eeq 
Assuming now that the brane width is of the order of inverse 
$\M4$, $B\sim 1/\M4$, and, furthermore, 
assuming that $\M4 >> M$, we arrive at the following estimate  for
the critical momentum and distance:
\beq
r^2_c~=~p_c^{-2}~\sim ~{\M4^N\over M^{2+N}}~.
\label{rpc}
\eeq
A simple estimate with $M\simeq $TeV and $\M4=10^{19}$ GeV
gives the following result for the $D=10$ case, $r_c\sim  10^{26}$ cm.
For smaller numbers of dimensions the value of $r_c$ decreases. 
At the distances smaller that $r_c$ we will observe 
the four-dimensional world. However, at  distances bigger then $r_c$ 
the laws of gravity would change\footnote{Note that these two
regimes should be matched at the distance scale $r_c$. This is non-trivial
task for gravitons, since the number of physical degrees of freedom 
for them can change. Thus, the matching for the case of a ``fat''
brane needs detailed investigations \cite {GGnew}.}.

\section{Comments  on  the Hierarchy Problem}

In the brane world scenarios with large compact
\cite {ADD} or warped extra dimensions with two branes 
\cite {RandallSundrum}
the hierarchy problem is solved due to the finite size 
of the extra space. The natural question is whether  the 
hierarchy problem could  be solved if the extra space has an infinite 
volume? Let us note that
the scale $M$ can in principle take any value  
between the 4D Planck scale, $10^{19}$ GeV,  and
the fundamental Planck scale of the framework \cite {ADD},  
that is a few TeV.  
In theories with infinite-volume extra dimensions there
is no phenomenological motivation for any particular value 
of $M$ in that domain. Thus, we keep $M$ as a free parameter. 
If $M$ is of order the 4D Planck scale, then 
there is no hierarchy    between $M$ and $\M4$ to explain.
However, one should still elucidate how the Higgs mass term is stabilized. 

On the other hand, if the scale $M$ is in a TeV region, 
then the big hierarchy between
$M$ and $\M4$ should somehow be justified.  
In fact, there are two issues to be addressed in this respect:
 
\begin{itemize}

\item{Given the high-dimensional fundamental scale $M$, how can we
obtain the scale on the brane, $\M4$,  
which is much bigger then $M$?}

\item{How does one stabilize the 
Standard Model Higgs mass against quadratically divergent 
radiative corrections?}

\end{itemize}

The answer to the first question is based on the  fact that
in theories  which allow  for brane solutions,
there  is a nonperturbative mass scale which is inversely proportional
to the coupling constant of the theory. If the coupling is very small
then the nonperturbative scale could be much bigger than the
fundamental scale of the model. Then, it is rather natural 
that the brane energy density (tension) is related to
the fundamental scale of the worldvolume theory as follows:
\beq
T~\propto~{M^4\over \lambda}~,  
\label{DT}
\eeq
where $\lambda$ is the coupling constant, or some other positive power
of the coupling constant  of the fundamental theory.
Let us ask the following question: What would be the value
of the induced Planck constant, $\M4$,
on the worldvolume of this brane. In this simplest setup 
this constant is determined
by the mass of the heaviest worldvolume particle that could propagate
in the loop which induces the 4D worldvolume Ricci scalar (see Appendix A).
Generically, if one  studies  quantum fluctuations of the brane
in the field theory context,
one  finds  that there is a heaviest fluctuation of the brane  with the mass
determined by the {\it inverse~ brane ~width}. 
Suppose  that for simplicity of arguments this scale is related to 
$M$ as follows (in general, the power of the coupling constant
in the expression below  can be different, however we would 
like to discuss a  
qualitative effect which is  independent of this assumption):  
\beq
\M4\propto (T)^{1/4}\propto {M\over \lambda^{1/4}}~.
\eeq
Thus, in the weak coupling approximation, when  $\lambda\to 0$,
we obtain  that
\beq
\M4~ >>~M~.
\eeq
In order for this hierarchy to be big, the coupling constant should be 
really tiny. Such a small coupling 
could not be a coupling of the worldvolume field theory
which should  have at least some resemblance to the Standard Model. 
In this respect, one  needs two sectors in the 
theory, one of them makes  the brane  with a huge tension 
as described above,  and another sector  governs  
the worldvolume physics. In fact, the situation might be a 
bit  more trickier.  
The relation (\ref {DT}) is 
exact for D-branes in string theory 
if one identified $M$ with the string scale and 
$\lambda$ with the string coupling constant. In this case,
the string coupling constant can be expressed via an exponent of 
the VEV of a dilaton
\beq
\lambda =\exp \left ( { <\phi> \over M} \right )~<<~1.
\eeq
Here we assume a certain mechanism for 
the dilaton stabilization which guarantees that the 
corresponding string theory is in a weak coupling regime.
One expects  that the field-theory cutoff of the brane worldvolume 
theory to be defined by $M$, since above this scale the string theory
description sets in. This is certainly true, and at scales above 
$M$ the higher stringy  modes should be taken into account. 
However, this does not eliminate the fact  that there  exist some
states which have masses that are much bigger that the string 
scale \cite {Shenker}. The simples example would be D0-branes with 
mass $m_{0}=M/\lambda $. The heavy  states could  also come from  
fluctuations of the brane itself. 
For instance, generically there is a localized massive mode on a solitonic
brane  the mass of which is determined by the inverse width of the 
brane. As we go below the energy scale 
determined by this  mass, the heavy state induces 
an appropriate worldvolume Ricci scalar. 
In this particular case, one can say that 
the 4D Ricci scalar is induced after 
integrating out the width of the brane. 
This will take place
regardless of the fact that other states (presumably perturbative 
string states) 
can be entering the problem at a lower scale $M$.  Thus, one way or other, 
the constant in front of the induced 4D Ricci term on the 
brane could  in principle be determined by this nonperturbative scale:
\beq
\M4~ \propto ~ M ~\exp \left (- { <\phi> \over 4M} \right )~>>~M.
\eeq
As long as the dilaton VEV is not a logarithmic function of 
a scale parameter, this gives rise to  an   {\it exponential} hierarchy 
between the 4D scale $\M4$ and the fundamental scale $M$.
For instance, the huge hierarchy between $M\sim$ (a few TeV) 
and $\M4 \sim 10^{19}$ GeV  could  be explained 
by a dilaton VEV which is only a $1/100$  part of the 
fundamental scale $M$. This latter relatively small hierarchy  
could in principle be obtained  as a result of 
some worldvolume coupling constant suppression of the corresponding terms. 
Note that as we discussed above, 
the coupling in the worldvolume theory
should not be determined by  the string coupling 
constant since this latter is extremely small. 
Thus, one needs again two sector in the fundamental theory:
One sector would create a brane with a big tension and another
one should be responsible for interactions in the brane worldvolume
theory. Something similar to the ``little string theory'' 
could do the job of the second sector \cite {WittenL}. 

Yet another possibility is to have a huge number of particles 
running in the loops which generate  the 4D Ricci scalar on the brane 
worldvolume\footnote{This possibility  was independently 
mentioned to us by D. Fursaev.}. In this case, $\M4$ can be 
much bigger than $M$ due to the multiplicity of states which induce 
$\R4$. This might be possible to realize in the constructions
where  the number of massive states increases polynomially or 
better yet exponentially. 
 
Let us now turn to the issue  of stabilization
of the Higgs mass  term. 
If there were SUSY on the brane, this would be the standard 
supersymmetric stabilization scenario.

If we deal with a  non-BPS brane, then 
there is  no supersymmetry on the worldvolume,
and the Higgs mass is driven by quadratically
divergent diagrams toward the cutoff of the worldvolume theory. 
If the cutoff of the low-energy field theory on a brane is 
$M\sim $ (a few TeV), then this would do the job 
of cutting the divergent diagrams and the Higgs mass would  
not need additional stabilization.   
On the other hand, if the cutoff is huge, 
then to avoid phenomenological difficulties, the Higgs field   
should be thought  of as a  composite field. The compositeness scale
can be somewhere in a few TeV region.  In particular, 
this composite Higgs could be coming from the properties of the bulk. 

Summarizing these discussions we emphasize  that 
there are possibilities  for solving the hierarchy  problem 
using the very  generic feature  of the brane  physics: 
In the presence of (D)branes 
there emerges a nonperturbative scale in the theory
which can be much bigger then the fundamental scale of a given model.
The detailed study of this  issue should be  
performed within the framework of concrete examples 
of branes and worldvolume field theories. 
This task is beyond the scope of the present work.

\section {On the Cosmological Constant problem}

The unnatural small value of the cosmological constant is a problem
shared by any theory of gravity which at low energies flows to
$D=4$  non-supersymmetric theory. Such is any
high-dimensional model  with broken supersymmetry and
{\it finite volume} extra space.

Apart from  a tree-level fine-tuning there is an important
issue of stability against quantum corrections. The  effective 4D
cosmological constant $\Lambda$ is power-law sensitive to the
cutoff of the theory  since it  gets renormalized 
by (at least quadratically) divergent loops. 

In a minimalistic scenario in which 
the low-energy theory is  Standard Model (SM) plus  Einstein's 
gravity (GR), one may conventionally split  the loop 
contributions (denote it by 
$\Delta\Lambda$)  to the 4D cosmological 
constant in the following two parts:
\beq
\Delta\Lambda ~= ~\Delta\Lambda_{\rm SM}~ + ~\Delta\Lambda_{\rm GR}~,
\label{split}
\eeq
where $\Delta \Lambda_{\rm SM}$ parameterizes the pure SM contribution,
whereas $\Delta\Lambda_{\rm GR}$ includes graviton loops.
Such a split  may be useful in theories with large extra dimensions
in which the SM particles are localized on a brane and gravity propagates
in the bulk. In this case  $\Delta\Lambda_{\rm SM}$ can be understood as
the renormalization of the brane tension by the SM loops.
In such a set-up $\Lambda$, in general, gets contributions
from both the brane tension $T$ and the bulk cosmological constant
$\Lambda_{\rm Bulk}$, so that we can write
\beq
 \Lambda ~= ~F(T,~\Lambda_{\rm Bulk})~,
\eeq
where $F$ is some model-dependent function. In particular,
 in the approximation
of flat $N$ compact extra dimensions one finds \cite{ADD}:
\beq
\Lambda =~ T~ + ~\Lambda_{\rm Bulk}~V_{\rm Extra}~,
\eeq
where $V_{\rm Extra}$ is the volume of extra space. 
For non-flat spaces the form of
$F$ may be more complicated. However, the net result is that in theories
with finite $V_{\rm Extra}$, the brane tension $T$ and $\Lambda_{\rm Bulk}$ 
must conspire with an
extraordinary accuracy in order to give $\Lambda \simeq 0$.
One may try to make $F$ insensitive to a brane tension
by introducing extra bulk degrees of freedom (perhaps coupled conformally
to the brane fields) \cite {ADKS,KS,Luty}. However, even if this is the case,
$\Delta\Lambda_{\rm GR}$ remains a big problem. One may expect naively that
graviton loops are  at least $1/M_{\rm Pl}^2$-suppressed, due to bulk SUSY.
However, this is not true
as it can be seen from the  following simple argument. 
The lowest scale at which
we have to break supersymmetry on a brane in the 
conventional approach is $\sim$ TeV. The Fermi-Bose
mass splitting
induced in the bulk by KK modes is then
\beq
 \Delta m^2 \sim {(\rm TeV)^4\over V_{\rm Extra}M^{2+N}}~.
\label{dm}
\eeq
Summing up one-loop contributions from all KK states lighter than $M$
and using the relation (\ref {dm})  we get
\beq
\Delta\Lambda_{\rm GR} \sim ({\rm TeV})^4~.
\label{problem}
\eeq
We cannot simply ignore this contribution, or attribute it to our
pure knowledge of quantum gravity, since it appears at a scale  lower than
$M$ where graviton loop contributions can be evaluated in effective
field theory. Quantum gravity theory sets in only above the scale 
$M$. While a miraculous cancellation  in  (\ref{problem}) 
{\it a priory} cannot be excluded,
it would imply some form of { \it non-decoupling}
of a very high energy physics for low energy observables.
This possibility will be disregarded in the present discussion.
To summarize, the finite volume theories face at least a potential
problem of radiative instability of the bulk cosmological constant.

Infinite volume theories, on the other hand,  
provide a loop-hole from the above argument
due to the fact that in these theories Eq. (\ref {dm}) is violated
\beq
{\M4^2 \over M^{2 + N}} \neq V_{\rm Extra}~ = ~\infty~.
\eeq    
As a result, even if supersymmetry is broken on a brane, it could 
still be preserved in the bulk. 
At all energy scales the theory remains to be a 
$(4 + N)$-dimensional supersymmetric model  (at least for $D> 6$).
The reason for this is that a  localized brane in $D>6$ 
gives an asymptotically flat metric in transverse directions.
On such a space there are infinitely many Killing spinors (any constant
spinor is a Killing spinors). This fact plays a crucial 
role in exact cancellation
of quantum gravity loops in renormalization of the
bulk cosmological constant.

Given the fact that we have the (super)symmetry reason for vanishing of
the bulk cosmological constant, it is time to ask what is the role
of the brane tension for an observable $\Lambda$. We shall argue now that
due to an infinite volume $\Lambda$ could  vanish for arbitrary $T$, provided
$D>6$ ($D=6$ case is also possible, but is somewhat subtle due to
the conical structure in the transverse space). 

Suppose there exists  a 3-brane solution to the D-dimensional 
Einstein equation:
\beq
M^{2+N} ~\left ({\cal R}_{AB}~-~{1\over 2}~g_{AB}~{\cal R}   \right )
~=~T_{AB}^{\rm Brane}~+T_{AB}^{\rm Other~fields}~,
\label{eom}
\eeq 
where $T_{AB}^{\rm Brane}$ is the brane energy-momentum tensor
and $T_{AB}^{\rm Other~fields}$ stands for the energy-momentum tensor
of other bulk and/or brane fields.

As long as $D\geq 6$ this equation can have 
a static solutions for a nonzero tension brane. 
That is to say, 
the four-dimensional worldvolume of the solution is flat. 
The line element for
these solutions takes the form:
\beq
ds^2~=~A(|y|)~{\bar g}_{\mu\nu}(x)dx^\mu dx^\nu~+~B(|y|) 
~\delta_{mn}~dy^m dy^n~,
\label{line}
\eeq
where $A$ and $B$ are the warp-factors which go  to 
constants at infinity. 
The statement that the solution has a flat four-dimensional 
worldvolume means  that 
\beq
{\bar g}^{\rm solution}_{\mu\nu}~=~\eta_{\mu\nu},~~~
\R4|_{\rm solution}~=~0~.
\label{eom1}
\eeq

Before proceeding further 
let us summarize  the main ingredients of the  framework:

 (I) The bulk cosmological constant should vanish  due to the 
bulk supersymmetry and R-symmetry (when the model is embedded in
SUGRA framework);
 
 (II) The brane worldvolume metric is flat, because in $D\geq 6$ spaces
there are flat brane solutions for an {\it arbitrary} tension $T$;

 (III) The induced curvature term $\R4$ on a brane worldvolume 
guarantees that gravity is four-dimensional on a brane.

To fulfill (III) we have to introduce 
the four-dimensional Ricci term $\R4$  on the worldvolume in Eq. 
(\ref {eom}).  The question is whether the 3-brane 
solution discussed  above still persists after the induced terms 
(\ref {4DR}) are taken  into account. We will argue that 
the answer is positive in the case at hand.
Indeed, 
with the induced terms included as in (\ref {4DR})
the Einstein equation of motion takes the form:
\beq
M^{2+N} ~\left ({\cal R}_{AB}~-~{1\over 2}~g_{AB}~{\cal R}   \right )~+~
\nonumber \\
\M4^2 ~\d ~A(0)~\left \{ -{1\over 2}  {\bar \Lambda} {\bar g}_{\mu\nu}~+~
~\left (\R4_{\mu\nu}~-~{1\over 2} {\bar g}_{\mu\nu} \R4    \right ) 
\right \} \delta ^\mu_A \delta^\nu_B ~=~\nonumber \\
T_{AB}^{\rm Brane}~+
T_{AB}^{\rm Other~fields}.
\label{eomind}
\eeq
Let us assume that $A(0)$ is a nonzero finite quantity, 
$A(0)<\infty$. 
Since on the solution with a  flat worldvolume $\R4=0$, then  
the induced terms in (\ref {eomind}) (all except the cosmological constant)
vanish on the solution.  The induced cosmological constant, 
on the other hand, can be re-absorbed into the brane tension
on the r.h.s. of the equation (\ref {eomind}). 
This rescaling of the tension changes  parameters of the
solution but not the form of the solution itself 
(similar to the change in the Schwarzschild solution  
caused by  the rescaling  of the mass of the spherically symmetric body). 
Therefore, we conclude that (if $A(0)<\infty$) 
equation (\ref {eomind}) also has a 
static solution which differs from the solution 
of Eq. (\ref {eom1})  by the redefinition of the brane tension:
\beq
T\to T~-~\M4^2 ~{\bar \Lambda}~ A (0)~.
\label{newT}
\eeq
Thus, we can obtain  a model  on  a brane 
which is embedded in an infinite-volume
extra space. This brane has zero 4D cosmological constant and the 
correct 4D relativistic worldvolume gravity. 

This, however, is not the solution to the cosmological problem yet. 
To really solve the problem, one should at least 
address the following three vital issues:
\begin{itemize}

\item {Why the solution with the flat brane worldvolume is unique?
In general there might be a number of other solutions
with an inflating worldvolume. 
As was suggested in  \cite {DGP1,WittenCC}, 
it could be  be bulk SUSY which might 
pick a single solution with a  flat worldvolume.} 

\item{Why is  the flat solution stable against phase transitions
on a brane worldvolume, let us say against the QCD or electroweak  
phase transitions?}

\item {Does the matter localized on a non-zero tension 
brane in the infinite-volume extra space  gives rise to
conventional (at least at some distances) Freedman-Robertson-Walker
cosmology?}

\end{itemize}

A 5D model was found in Ref. \cite {Z2} in which
bulk supersymmetry does control  the worldvolume cosmological constant
on a brane. On the other hand, the model is 
very restrictive so that the worldvolume theory turns out to 
be conformal \cite {Z3}. 

Whether these issues can be answered positively in the present 
$D>5$ framework (along with the issue of the existence of 
different solutions) is the subject of an ongoing investigation   
and will be reported elsewhere.

\section{Discussions and conclusions}

The main objective  of the present work was to
explore the possibility of generating 
a relativistic  4D theory of gravitation on a 
singular brane which is embedded in a flat infinite-volume extra space.
Four-dimensional  gravity is obtained on a brane worldvolume 
due to the induced 4D Ricci scalar \cite {DGPind}.
We showed  that if the number of extra dimensions is
two or bigger, then the induced gravity on a {\it delta-function}
type brane is a four-dimensional relativistic tensor theory. 
In particular, the tensor structure of the graviton propagator in this
case is equivalent  to that of Einstein's gravity.

In the present framework the extra  space exhibits 
the phenomenon of infrared transparency. 
That is to say, the bulk  can only be probed by the signals with 
zero 4D momentum square. 

This approach offers 
new opportunities  to study  the cosmological constant problem. 
In particular, the bulk cosmological constant in this framework 
can be controlled 
by the bulk symmetries, while the brane cosmological constant can be 
re-absorbed into  the brane tension. The generalization for
the case of a ``fat'' brane and the question whether this framework could  
lead to a unique brane solution with a flat 4D worldvolume 
will be discussed elsewhere.

\vspace{0.4cm}

{\bf Acknowledgments}
\vspace{0.1cm} \\

We would like to thank B. Bajc, S. Dimopoulos,
V. Dubrovich, D. Fursaev, Z. Kakushadze, N. Kaloper, E. Poppitz, 
M. Porrati, V. Rubakov, M. Voloshin 
and R. Wagoner for useful discussions. 
The work of G.D. is supported in part by a David and Lucile  
Packard Foundation Fellowship for  Science and Engineering. G.G. is
supported by  NSF grant PHY-99-96173. 

\newpage

\vspace{0.7in} 

{\large \bf Appendix A}

\vspace{0.2in}

In this Appendix  we summarize the mechanism  by which 
expression (\ref {4DR}) can emerge on a brane worldvolume 
\cite {DGPind}. 
It is instructive to consider two different possibilities.
If the brane is ``rigid'' then 
quantum fluctuations of the matter 
fields which are localized on the  
brane can induce the term 
(\ref {4DR}) via loop diagrams. 
On the other hand, if the brane is allowed to fluctuate 
then generically there is at least one  massive state
in the spectrum of fluctuations of this brane.
This state  could for instance correspond  to 
the breathing mode of the transverse size of the brane. 
From the point of view of a worldvolume observer
this state looks  as a massive mode  of the 
worldvolume theory. 
Therefore, this mode  can also run in 
loops and produce the terms (\ref {4DR}).  
Below, we present our discussion in 
terms of the states of the 
localized matter on the brane. However, we keep in mind that the very 
same consideration applies to the massive fluctuations of the brane 
itself. Having this said, we write the matter energy-momentum tensor 
as follows: 
\beq
T_{AB}~=~\left (
\begin{tabular} {c c}
$T_{\mu\nu}(x)~ \delta^{(N)} (y_m)$ & ~~0 \\
0              & ~~0    \\
\end{tabular}
\right )~.
\nonumber 
\eeq
As a result, the interaction Lagrangian of localized 
matter with  D-dimensional  
metric fluctuations $h_{AB}(x,y)\equiv G_{AB}(x,y)-\eta_{AB}$, 
reduces to the following expression: 
\beq
{\cal L}_{\rm int} ~=~ \int d^Ny~h^{\mu\nu}(x,y_m) 
~T_{\mu\nu}(x)~\delta^{(N)}(y_m)~=~
h^{\mu\nu}(x,0) ~T_{\mu\nu}(x)~,
\nonumber
\eeq
where the 4D induced metric 
${\bar g}_{\mu\nu}(x)=\eta_{\mu\nu}+h_{\mu\nu}$ 
was  defined earlier  in (\ref {gind}).
Due to this interaction, a 
4D kinetic term can be generated  for 
${\bar g}_{\mu\nu}(x)$  in the full quantum theory. 
For instance, the diagram  of Fig. 1 with massive 
scalars \cite {Capper},  or fermions \cite {Adler,Zee}
running in the loop  would  induce the following  4D term
in the low-energy action:
\beq
\int d^4x~d^Ny~\delta^{(N)}(y_m)~\sqrt{|g|}~\R4~.
\nonumber
\eeq 
\begin{figure}
\centerline{\epsfbox{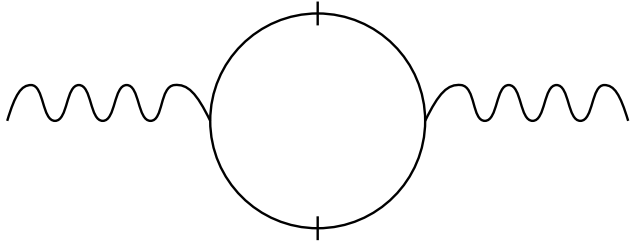}}
\epsfysize=6cm
\caption{\small The  one-loop diagram which generates  the 
4D Ricci scalar $\R4$. Wave lines denote gravitons, 
solid lines denote massive scalars/fermions. 
Vertical short lines on 
scalar/fermion propagators indicate that they are massive.}
\label{fig1}
\end{figure} 
The corresponding induced gravitational constant will be determined by
a correlation function of the world-volume matter theory.
The magnitude of this constant, as we have discussed in the text, 
depends on a worldvolume theory at hand and is vanishing  
in conformally invariant models, or nonzero 
if conformal invariance is broken 
(for detailed discussions see \cite {Adler,Zee,Khuri}).
We assume that
the worldvolume theory is not conformal 
and the second term in (\ref {1}) is generated.

As we discussed,  in general one induces on a brane 
the whole series in powers of the four-dimensional
Ricci scalar $\R4$. The very first term in this 
series is the induced 4D cosmological constant, 
${\bar \Lambda} =\langle 0|T^{\mu}_{\mu}|0\rangle$.  
The higher-derivative terms can also be generated.
We have to deal with these contributions separately. Let us start with 
the induced four-dimensional cosmological constant.
As we discussed in section 6, this just renormalizes  the 
brane tension in $D>5$, and does not change the worldvolume physics
when the static brane solution exists. 

Let us turn to the higher derivative terms. These are suppressed 
by higher powers of $\M4$, thus their effects on 4D worldvolume 
gravity should be small. The only subtlety with 
these terms is that in certain  cases they can 
give rise to ghosts in 4D  theory. 
However, these ghosts are absent if the corresponding 
high-derivative terms come in certain combinations. 
For instance, it is known that in the 
second  order in $\R4$ the  ghosts are absent  if the $\R4^2$
terms come in the Gauss-Bonnet combination \cite {Zwiebach}. 
We will be assuming that the bulk and worldvolume theories   are
consistent models  with no propagating ghosts or other unconventional 
states. Therefore, the resulting expression
for induced Lagrangian  on the brane is expected to be 
ghost free in such theories\footnote{It might also happen that
the bulk theory is ghost free, however   ghosts emerge as artifacts  
of the truncation of the perturbative series in $\R4$. In this case 
the perturbative approach with consistent subtraction
schemes in each order of perturbation theory should be developed.}.

\vspace{0.2in}

{\large \bf Appendix B}

\vspace{0.2in}

In this Appendix we recall the properties of the 5D 
theory. We follow the discussions in \cite {DGPind}.
At the end of the appendix we discus some novel features.

The crucial  difference from the theories in higher co-dimensions is that
the Green function in the transverse space in 5D is finite at the origin.
In the notations of Eq. (\ref {ED}):
\beq
D(p,y)~=~{1\over 2p}~\exp \{-p|y|\}~.
\nonumber
\eeq
As a result, the expression for the retarded Green function 
(\ref {G}) takes the form \cite {DGPind}:
\beq
{\tilde G}_R(p,y)~=~{1\over 2M^{3}p~+~\M4^2p^2}~\exp\{-p|y|\}~.
\nonumber 
\eeq
This gives rise to the potential which has
4D behavior at observable distances, but 5D behavior
at ultra-large scales \cite {DGPind}. In this model, as we mentioned 
above, very low energy gravitons can leak into the bulk space
even when the brane width is zero \cite {GRS,Csaki1,DGP1}.

Let us now study the tensor structure of the graviton propagator
\cite {DGPind}. The ($\mu\nu$) components
of the Einstein equation take  the form:
\beq
\left ( M^3 \partial_A\partial^A~+~\M4^2 ~\delta(y)~\partial_\mu\partial^\mu 
\right )~h_{\mu\nu}(x,y) ~=~ \left \{ T_{\mu\nu} -{1\over 3}\eta_{\mu\nu} 
T^\alpha_\alpha   \right \}~\delta(y)~\nonumber \\
+~ \M4^2~\delta(y) ~\partial_\mu
\partial_\nu ~h^5_5~.  
\nonumber
\eeq
This has a structure of a massive 4D graviton or, equivalently 
that of a massless 5D graviton. 
In this respect, it is instructive to rewrite
this  expression  in the following form:
\beq
\left ( M^3 \partial_A\partial^A~+~{\M4}^2~\delta(y)~
\partial_\mu\partial^\mu 
\right )~h_{\mu\nu}(x,y) ~=~ \left \{ T_{\mu\nu} -{1\over 2}\eta_{\mu\nu}  
T^\alpha_\alpha   \right \}~\delta(y)~ \nonumber \\
-{1\over 2} 
~M^3 ~\eta_{\mu\nu} ~\partial_A\partial^A~h^\alpha_\alpha
~+~ \M4^2~\delta(y)~ \partial_\mu \partial_\nu ~h^5_5~.  
\nonumber
\eeq
Here the tensor structure on the r.h.s. is that
of a 4D massless graviton. However, there is an additional
contribution due to the trace part $h^\mu_\mu$ which is nonzero.  
Therefore, one is left with 
the theory of gravity which from the 4D point of view is mediated
by a graviton plus a scalar.
As before, turning to the Fourier images in the Euclidean space
we find:
\beq
{\tilde h}_{\mu\nu}(p, y=0)~{\tilde T}^{\mu\nu}(p)~=~
{  {\tilde T}^{\mu\nu}{\tilde T}_{\mu\nu}~-~{1\over 3} ~{\tilde T}^{\mu}_\mu
{\tilde T}^{\nu}_\nu \over \M4^2 p^2~+~2M^3 p }~.
\nonumber
\eeq  
Here the tilde sign  denotes  the Fourier-transformed
quantities. 
Thus, the tensor structure of the  graviton propagator in 4D worldvolume 
theory  looks as follows:
\beq
{1\over 2} \eta^{\mu\alpha}  \eta^{\nu\beta}~+~
{1\over 2}  \eta^{\mu\beta}  \eta^{\nu\alpha}~-~   
{1\over 3}  \eta^{\mu\nu}  \eta^{\alpha\beta} ~+~{\cal O} (p).
\nonumber
\eeq
At short distances the potential scales as 
$1/r$ with the logarithmic corrections found in \cite {DGPind}. 
On the other hand, at   
large distances  the $1/r^2$ 
behavior is recovered.
The tensor structure of the propagator is that 
of 4D tensor-scalar gravity.  

The presence of the extra polarization degrees of freedom is
{\it not} acceptable from the phenomenological point of view
\cite {Veltman,Zakharov}. The light bending by the Sun and the 
precession of the Mercury perihelion in this theory are 
incompatible with the existing data.
The reason for this is that gravity in these models is mediated
not only by two transverse polarizations of a 4D graviton, but 
also, by an additional polarization  of a 5D-dimensional graviton. 
Thus, there is the excess of attraction in the theory. 
This extra attraction should  be somehow removed in order 
to render the model compatible with the data.

A  way to  avoid  the problem with the 
extra degrees of freedom in the worldvolume theory is 
to assume that the bulk gravity is 
in fact described by a topological theory
which has no propagating degrees of freedom. 
As an example one could  write in 5D the 
Chern-Simons term \cite {Zumino}.
The terms in  (\ref {4DR}) will still be induced 
in the worldvolume theory as long  as 
the brane is introduced (this is true for any number of dimensions). 
Therefore, the only 
propagating degrees of freedom will be those of 
4D gravity on a brane worldvolume. 
These latter appear due to  (\ref {4DR}). 
In terms of equations this could be seen as follows:
in a model with the bulk Chern-Simons term the first  
term in Eq. (\ref {basic}) is  absent \cite {Zumino}. Thus,
the tensor structure of the graviton propagator 
is  determined  by 
the second term in (\ref {basic}). Hence, 
the resulting tensor structure is  
that of four-dimensional gravity:
\beq
{1\over 2} \eta^{\mu\alpha}  \eta^{\nu\beta}~+~
{1\over 2}  \eta^{\mu\beta}  \eta^{\nu\alpha}~-~   
{1\over 2}  \eta^{\mu\nu}  \eta^{\alpha\beta} ~.
\nonumber
\eeq
Let us note that higher dimensional 
topological gravity could  in particular be obtained  
from certain  compactifications of string theory 
\cite {Vafa}\footnote{We thank Zurab Kakushadze for
stimulating discussions on these  issues.}. 
The net result  of this approach is that the 
worldvolume gravity on a brane is four-dimensional 
to  all distances. That is to say, there is no 
crossover to the high-dimensional gravitational law at
large distances.  The theory in this case is intrinsically 
four-dimensional, that is to say no localized matter
can escape from the brane to the bulk.

Finally one could  incorporate scalars into the consideration
in co-dimension-one theories \cite {Zbig,Z2,Z3}.

\end{document}